\documentclass[reprint,
superscriptaddress,
 amsmath,amssymb,
 aps,
pra,
]{revtex4-1}

\usepackage[pdftex]{graphicx}
\usepackage{dcolumn}
\usepackage{bm}
\usepackage{amssymb}
\usepackage{amsmath}
\usepackage{color}

\definecolor{red}{rgb}{1,0,0}

\newcommand{\ket}[1]{|#1\rangle}
\newcommand{\bra}[1]{\langle #1|}

\begin{document}

\title{Extreme non-linear response of ultra-narrow optical transitions
  in cavity QED for laser stabilization}

\author{M. J. Martin}
\author{D. Meiser}
\affiliation{JILA, National
  Institute of Standards and Technology, and Department of Physics,
  The University of Colorado, Boulder, Colorado 80309-0440, USA}
\author{J. W. Thomsen}
\affiliation{The Niels Bohr Institute,
  Universitetsparken 5, 2100 Copenhagen, Denmark}
\author{Jun Ye}
\author{M. J. Holland}
\affiliation{JILA, National Institute of Standards
  and Technology, and Department of Physics, The University of
  Colorado, Boulder, Colorado 80309-0440, USA}

\date{\today}



\begin{abstract}
  \noindent We explore the potential of direct spectroscopy of ultra-narrow
  optical transitions of atoms localized in an optical cavity.  In contrast to
  stabilization against a reference cavity, which is the approach
  currently used for the most highly stabilized lasers, stabilization
  against an atomic transition does not suffer from Brownian thermal noise.  Spectroscopy of ultra-narrow optical transitions in a
  cavity operates in a very highly saturated regime in which
  non-linear effects such as bistability play an important role. From the universal behavior of the Jaynes-Cummings model with dissipation, we derive  the fundamental limits for laser stabilization using direct spectroscopy of ultra-narrow atomic lines.  We find that with
  current lattice clock experiments, laser linewidths of about 1~mHz can
  be achieved in principle, and the ultimate limitations of this technique are at 
  the 1 $\mu$~Hz level.
\end{abstract}

\maketitle

\noindent Ultra-stable lasers are central components of optical atomic clocks
and precision spectroscopy.  Today's most stable lasers are made by
locking the frequency of a prestabilized laser to a resonance of a
high finesse reference
cavity~\cite{Drever:FrequencyStabilization, young99, Jiang2011}.
The phase stability of these lasers is limited by thermal noise in the
mirrors of the reference cavity \cite{numata04}.  They achieve linewidths below 1
Hz~\cite{Ludlow:CavityStabilization} corresponding to oscillator
quality factors ($Q$-factors) of order $10^{15}$. Improving laser stability beyond the current state of the art will have a significant impact on precision science and quantum metrology \cite{diddams2004standards}, but further advances in laser stability through refinement of reference cavities requires a significant investment in resources given the
maturity of the optical designs involved \cite{chen06}.  The purpose of this paper is to propose an alternative laser stabilization technique, by means of direct cavity-enhanced nonlinear spectroscopy, and to elucidate the rich phenomenology of this approach in an extreme regime of cavity quantum electrodynamics and optical bistability.

Strong optical transitions typically used for laser stabilization
 are not suitable for ultimate laser stability since the atomic transition frequency is very
sensitive to stray fields, collisions, etc.  However, for special
ultra-narrow optical clock transitions that are now being routinely
used for optical atomic
clocks~\cite{T.Rosenband03282008EtAl,ADLudlow03282008EtAl, lemke2009}, these
shifts are small, very well characterized, and can in some cases be
eliminated or controlled~\cite{Swallows:FermionicCollisions}.

Compared to the use of strong transitions, the physics of this
frequency locking scheme is non-trivial because the atomic transition
is strongly saturated for very small intensities.  Additionally, sufficient free-space optical depths are not available in current-generation experiments. One can circumvent this problem by working in a cavity-enhanced, highly non-linear, strongly saturated regime in order to
achieve a signal that is strong enough for laser feedback.
This regime has been studied extensively in the context of non-linear
optics with alkali
atoms~\cite{Drummond:Bistability,Gripp:AnharmonicityRabiPeaks,Mielke:TimeResponse,Foster:IntensityCorrelations}, albeit in a much less extreme limit.

\begin{figure}
\begin{center}
\includegraphics[clip=true,width=7cm]{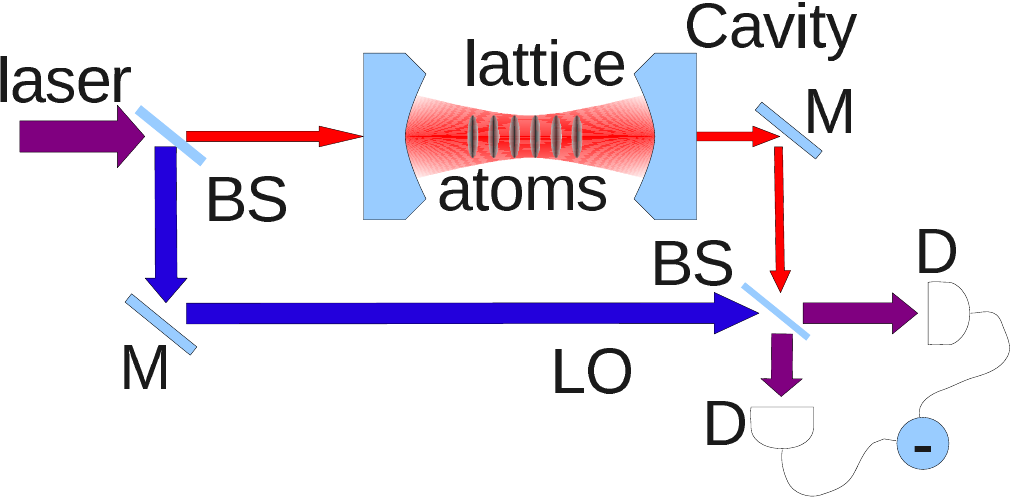}
\end{center}
\caption{(Color online) Schematic of cavity enhanced ultra-narrow
  linewidth absorption spectroscopy for laser stabilization, M:
  mirror; BS: beam splitter; LO local oscillator; D: photodiode.}
\label{fig:AtomCavitySchematic}
\end{figure}

In this paper we consider a simplified model that contains all the
essential components of this many-atom cavity QED system
(Fig. \ref{fig:AtomCavitySchematic}), but in the extreme bad cavity limit. Here, despite the unavoidable strong saturation effect, we are able to uncover a collective atomic interaction regime where we preserve the superior frequency discrimination capability of a narrow atomic transition.    
This model serves as a basis upon which to calculate the fundamental
limitations of our stabilization scheme, although real-world implementations will require more
complicated topologies.  One such approach could be based on the NICE-OHMS technique \cite{Ye1998, Foltynowicz2008},where the local oscillator and signal beams are co-propagated through the cavity to reject common-mode frequency noise. The effects of finite vacuum lifetime and heating could be addressed by operating two systems in a multiplexed fashion, while heating could additionally be mitigated at the single-system level by implementing a Raman cooling scheme similar to that proposed in \cite{Meiser:SrLaser}.

In our simplified theoretical analysis, we consider an ensemble of $N$ two-level
atoms with transition frequency $\omega_a$ trapped in an optical
lattice potential inside a cavity.  The lattice is at the magic
wavelength where the difference of the AC Stark shifts of both levels
vanish \cite{ye2008}.  The atoms are assumed to be in the vibrational ground state
along the lattice direction and in the Lamb-Dicke regime such that we
can neglect Doppler broadening and recoil effects.  The atomic transition is near
resonant with a cavity resonance with frequency $\omega_c$ and
field decay rate $\kappa$.  A laser with frequency $\omega_L$ is coupled into
the cavity and the transmitted light is detected by means of
balanced homodyne detection.

This $N$-atom system is described by the Hamiltonian
\begin{eqnarray}
\hat H &=&
\frac{\hbar \omega_a}{2}\sum_{j=1}^N\hat \sigma_z^{(j)} +
\hbar \omega_c\hat a^\dagger \hat a +
\hbar \eta \left(e^{-i\omega_L t}\hat a^\dagger+\text{h.c.}\right)\nonumber\\*
&& + \hbar g\sum_{j=1}^N\left(\hat a^\dagger \hat \sigma_-^{(j)}
+ \text{h.c.}\right)\;.
\label{eqn:Hamiltonian}
\end{eqnarray}
Here, $\hat \sigma_z^{(j)}=\ket{e_j}\bra{e_j}-\ket{g_j}\bra{g_j}$ is the operator
for the inversion of atom $j$, and $\hat
\sigma_+^{(j)}=\ket{e_j}\bra{g_j}$ and $\hat
\sigma_-^{(j)}=\ket{g_j}\bra{e_j}$ are spin raising and lowering
operators, respectively.  The bosonic field operator
$\hat a$ is the annihilation operator for a photon in the cavity.  The
coupling constant $g= \left(\wp/\hbar\right) \sqrt{\hbar \omega_c/\left(2V_{\rm
    eff}\epsilon_0\right)}$ is half the vacuum Rabi frequency with
$V_{\rm eff}$ the effective mode volume of the cavity, $\wp$ the
dipole moment of the atomic transition, and $\epsilon_0$ the vacuum
permittivity.  The cavity is classically driven with amplitude $\eta$ by the
in-coupled laser. 

In addition to the coherent dynamics described by the Hamiltonian we
also need to account for dissipative processes.  These are spontaneous
emission from the excited atomic state (decay rate
$\gamma$), decay of the atomic dipole with rate
$T_{2}^{-1}$, and decay of the cavity field with rate $\kappa$.  We
treat these dissipative processes within the usual Born-Markov master
equation \cite{walls2008}.  Although we do not consider inhomogeneous atom-cavity coupling, this effect does not change our results qualitatively and can in principle be taken into account primarily by a rescaling of the cooperativity parameter via an effective atom number.

We assume that the cavity is locked to the probe laser, i.e., $\omega_L=\omega_c$.  This could be achieved for example
by using a frequency-offset Pound-Drever-Hall locking scheme \cite{drever1983} on a different cavity
longitudinal mode in conjunction with a piezo-tuneable cavity.  Effects due to a
slight detuning between laser and cavity are negligible owing to the
comparatively large cavity linewidth and we further quantify this statement in Appendix~\ref{LinePull}.

To study the non-linear dynamics of this system we consider a
semi-classical approximation where all expectation values of more than
one operator can be factorized, e.g. $\langle\hat a^\dagger \hat
\sigma_-^{(j)}\rangle\approx \langle \hat a^\dagger \rangle \langle
\hat \sigma_-^{(j)}\rangle$.  Consequently, we find the set of first order equations of
motion for the expectation values $o\equiv\langle\hat o\rangle$ with
$\hat o \in \{ \hat a\,, \hat \sigma_-\,, \hat\sigma_z\}$,
\begin{eqnarray}
\frac{d a}{dt} &=& \eta-\kappa a + gN\sigma_-\label{eqn:Amplitude}\\
\frac{d \sigma_-}{dt} &=& - (T_{2}^{-1} + i \Delta) \sigma_- + g a \sigma_z\label{eqn:Dipole}\\
\frac{d \sigma_z}{dt} &=& - \gamma (1+\sigma_z)-4g{\rm Re}(a \sigma_-^*)\;.\label{eqn:Inversion}
\end{eqnarray}
The atom--cavity detuning is $\Delta = \omega_a -
\omega_c$.

The steady state of the system is obtained by setting the time
derivatives to zero.  The steady state polarization of the atoms is
given by
\begin{equation}
\sigma_- = \frac{g}{T_{2}^{-1}+i\Delta}a\sigma_z\;.
\end{equation}
Inserting this into the equations for the inversion we find the
saturated inversion
\begin{equation}
\sigma_z=\frac{-1}{1+\frac{|a|^2/n_0}{1+T_{2}^{2}\Delta^2}}\;,
\end{equation}
where $n_0=\gamma T_{2}^{-1}/(4g^2)$ is the saturation
photon number.  The mean number of photons in the cavity is then
\begin{equation}
  |a|^2=\frac{\eta^2}{\kappa^2}
  \frac{1+T_{2}^{2}\Delta^2}{(1-\mathcal{C}\sigma_z)^2+T_{2}^{2}\Delta^2}\;.
\end{equation}
Here, $\mathcal{C}=N\mathcal{C}_0$ is the cooperativity
parameter and $\mathcal{C}_0=g^2/(\kappa T_{2}^{-1})$ is the single
atom cooperativity parameter.

\begin{figure}
\begin{center}
\includegraphics[trim = 0mm 0mm 0mm 0mm, clip]{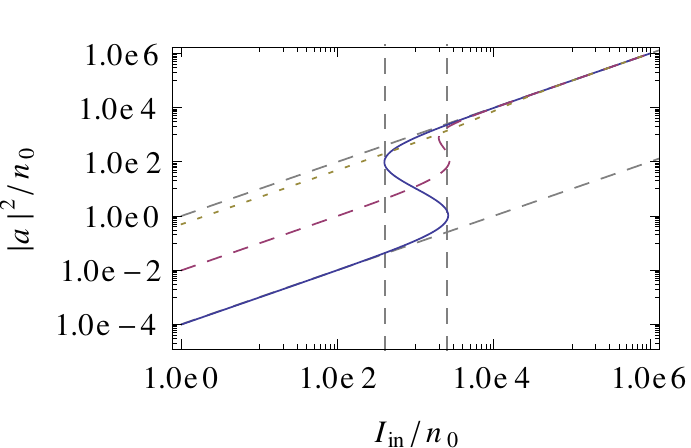}
\end{center}
\caption{(Color online) Intra-cavity intensity as a function of
  incoupled intensity for $\mathcal{C}=100$ and $\Delta = 0$ (blue
  solid line), $\Delta = 10 T_{2}^{-1}$ (purple dashed line), and
  $\Delta = 100 T_{2}^{-1}$ (yellow dotted line).  The vertical
  dashed lines mark the lower and upper threshold for bistability.
  The diagonal dashed lines show the intra-cavity intensity for a
  completely bleached atomic ensemble, $|a|^2=I_{\rm in}$, and for the
  unsaturated limit, $|a|^2=I_{\rm in}/(2\mathcal{C})^2$.  The
  saturation photon number is
  $n_0=\gamma T_{2}^{-1}/(4g^2)$.}
\label{fig:BistablityCurve}
\end{figure}

In this proposal we consider a regime of high-cooperativity where the total optical depth of the atom-cavity ensemble is greater than unity in the weak-driving limit. Specifically, in order to enter the nonlinear regime of spectroscopy considered here, the total cooperativity must satisfy $\mathcal{C}>8$.  
The solution for the steady state intensity with $\mathcal{C}=100$ is illustrated in
Fig.~\ref{fig:BistablityCurve}.  For low in-coupled intensity, $I_{\rm
  in} \equiv \eta^2 /\left(n_{0}\kappa^2\right)<4\mathcal{C}$, the atoms and cavity
behave like two coupled harmonic oscillators.  For $\omega_a=\omega_c$
the resonances of the coupled system are split by $2 g\sqrt{N}$, the vacuum Rabi
splitting.  Hence, the driving field is far detuned from the
coupled-system resonances for $\Delta = 0$ and the intensity inside
the cavity is reduced by a factor $1/\mathcal{C}^2$ compared to an
empty cavity.  On the other hand, in the strong driving limit, $I_{\rm
  in }>\mathcal{C}^2/4$, the atomic transition is completely saturated
and the cavity behaves as if it were empty.  In the intermediate
regime, $4\mathcal{C}<I_{\rm in}<\mathcal{C}^2/4$, two stable solutions
exist; a low intensity branch on which the atomic transition is
unsaturated and a high intensity branch on which the atomic transition
is saturated.

\begin{figure}
\begin{center}
\includegraphics[trim = 0mm 5mm 0mm 5mm]{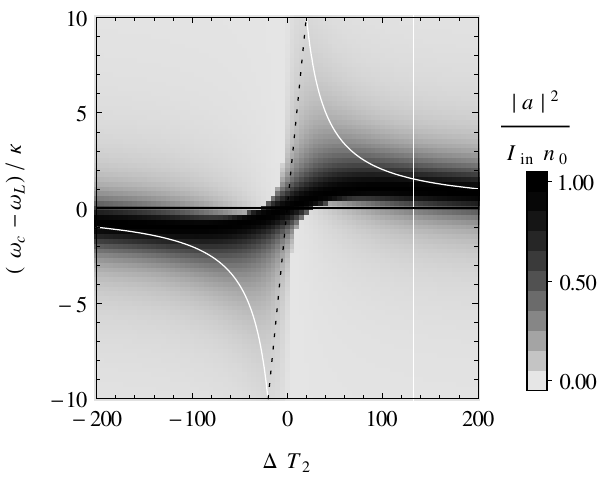}
\end{center}
\caption{Intra-cavity intensity as a function of detuning of the
  driving laser from the atomic resonance and from the cavity for
  $\mathcal{C}=100$ and $I_{\rm in} = 5\times 10^3$.  Only the
  solution with the largest intra-cavity intensity is shown.  Near
  resonance there are two additional solutions (see
  Fig. \ref{fig:NonLinearSpectra}).  The white hyperbolas indicate the
  resonances of the weakly driven system. 
  }
\label{fig:SurfacePlots}
\end{figure}
To clarify the connection of the physics considered here with previous
studies of optical bistability in cavity QED, it is useful to consider
the intensity in the cavity as a function of $\Delta$ and
$\omega_c-\omega_L$.  One of the stable solutions for the intra-cavity
intensity is shown in Fig. \ref{fig:SurfacePlots}.  In the weak
driving limit, $I_{\rm in}\to 0$, the resonances of the system
approach the white hyperbolas while the resonance of the strongly
driven system, $I_{\rm in}\to\infty$, lies on the black horizontal
line.  Remarkably, with the axis rescaled as in that figure, the plots
depend only on two free parameters, $C$ and $I_{\rm in}$.  Most
experiments on optical bistability in cavity QED to date have been
carried out in a regime where $C/T_{2} \gg \kappa$.  For such an
experiment, scanning $\omega_L$ with $\omega_a=\omega_c$ corresponds
to the nearly vertical dotted line in this
figure~\cite{Gripp:AnharmonicityRabiPeaks,Gripp:EvolutionRabiPeaks}.
In our proposal $\Delta$ is scanned while $\omega_c=\omega_L$ at all
times, corresponding to the black horizontal line.  While the basic physics
behind this non-linear coupled system has been known for a long
time~\cite{Gripp:EvolutionRabiPeaks}, it has not been interrogated in
the way discussed here.

\begin{figure}
\begin{center}
\includegraphics[trim = 0mm 0mm 0mm 0mm, clip]{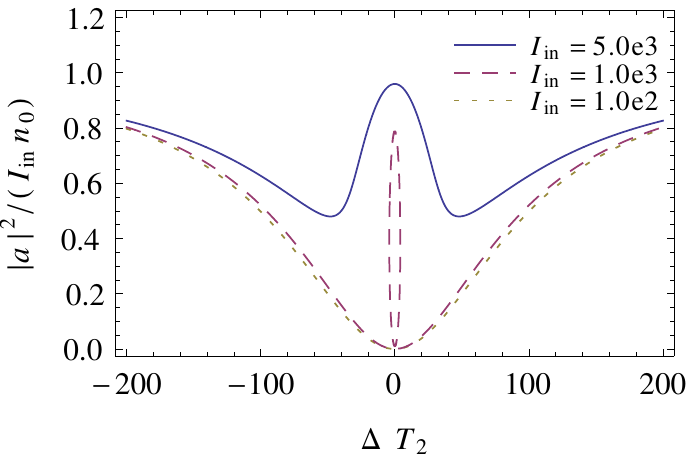}
\end{center}
\caption{(Color online) Intra-cavity intensity for $\mathcal{C} = 100$ as a function of
  detuning for incoupled intensities $I_{\rm in}=5\times 10^3$
  (blue solid line), $I_{\rm in}=1\times 10^3$ (purple dashed line), and
  $I_{\rm in}=1 \times 10^2$ (yellow dotted line).}
\label{fig:NonLinearSpectra}
\end{figure}

The spectra resulting from scanning $\Delta$ in this way are shown for
weak, intermediate (i.e., bistable), and strong pumping in
Fig. \ref{fig:NonLinearSpectra}.  These spectra are cuts through the
plot in Fig.~\ref{fig:SurfacePlots} along the $\Delta=0$ line.  In the
weak pumping regime (dotted line) we see a broadened absorption
feature with width $\mathcal{C}T_{2}^{-1}$. In the bistable regime
(dashed line) there are three possible stationary values of the
intra-cavity intensity near resonance.  The solutions corresponding to
largest and smallest intensity are dynamically stable while the
intermediate intensity solution is dynamically unstable.  In the
strong pumping regime (solid line) there is only one steady state for
any detuning and a \emph{peak} develops near resonance. Physically,
this peak emerges because near resonance the atomic transition is strongly saturated, whereas away from resonance the cavity field experiences
an additional phase shift due to the atoms and does not build up in the cavity.

\begin{table*}[]
\begin{tabular}{|l||c|c|c|c|c|c||c|c|c|}
\hline
Transition&
$\lambda$&
$T_{2}^{-1}$&
$\gamma$&
$N$&
$\mathcal{F}$&
$\mathcal{C}_0$&
$P(\beta=2)$&
$\mathrm{SNR}$&
$\Delta\nu$\\
\hline
\hline
$^{24}$Mg ${}^1S_0\rightarrow{}^3P_1$&
457 nm&
$\gamma/2$&
$2\pi\times 31$ Hz&
$10^4$&
$10^4$&
$9.6\times10^{-3}$&
20 pW&
$9.8 \times 10^{3} $&
20 mHz
\\
${}^{87}$Sr ${}^1S_0\rightarrow{}^3P_0$&
698 nm &
$1$ s$^{-1}$&
$2\pi\times 1$ mHz&
$10^5$&
$10^5$&
$7.4\times 10^{-4}$&
$3$ fW&
$1.5\times 10^{2}$& 
$4.7$ mHz\\
${}^{171}$Yb ${}^1S_0\rightarrow{}^3P_0$&
578 nm &
$1$ s$^{-1}$&
$2\pi\times 44$ mHz&
$10^4$&
$5\times 10^4$&
$1.1\times 10^{-2}$&
$27$ fW&
$3.9\times 10^{2}$& 
$1.6$ mHz\\
${}^{199}$Hg ${}^1S_0\rightarrow{}^3P_0$&
265.6 nm &
$1$ s$^{-1}$&
$2\pi\times 100$ mHz&
$10^4$&
$10^5$&
$1.1\times 10^{-2}$&
$130$ fW&
$5.8\times 10^{2}$& 
$0.68$ mHz\\
${}^{87}$Sr ${}^1S_0\rightarrow{}^3P_0$&
698 nm &
$\gamma/2$&
$2\pi\times 1$ mHz&
$10^4$&
$5 \times 10^3$&
$1.2 \times 10^{-2}$&
$0.5$ fW&
$6.1\times 10^{1}$&
0.74 $\mu$Hz\\
\hline
\end{tabular}
\caption{Quantum limited linewidth according to
  Eq. (\ref{eqn:linewidth}) for several optical lattice clock systems.
  The cavity geometry is $V_{\rm eff}=L \times (100\,
  \mu{\rm m})^2$ and the finesse, $\mathcal{F}$, is tuned to give
  $N\mathcal{C}_0\simeq100$.  The signal to noise ratio (SNR) is 1 Hz bandwidth-normalized.  In all but the last case, $T_{2}$ values have been set to be $\leq$1~s. This is a conservative estimate based on current-generation lattice clock experiments  \cite{Boyd2006}. }
\label{tab:SpecificSystems}
\end{table*}

In this work, the new idea is to lock the probe laser and cavity to this strongly
saturated resonance feature.  To estimate the potential performance of
such a lock, we need to know the signal power and the slope of the
phase across the resonance.  The signal power is equal to the power leaking out of the cavity in steady-state, and is given by
\begin{equation}
P\simeq \hbar \omega_L \kappa \mathcal{C}^2 n_0 \beta/2 = 2 \hbar \omega_{L}  \eta^{2}/\kappa \;,
\label{Power}
\end{equation}
where the parameter $\beta=4 I_{\rm in}/(\mathcal{C}^2)\gtrsim 1$
describes how far above the upper threshold for bistability the system
is driven.  This power corresponds to a photon shot noise limited
bandwidth-normalized signal to noise ratio of ${\rm SNR}^{2} =\kappa
  \mathcal{C}^2n_0\beta \; \left[\mathrm{Hz} \right]$, assuming unity photodetector quantum efficiency.  Specifically, near resonance, we can write the differential photocurrent from the system as
  \begin{equation}
  i_{\mathrm{diff}} = \frac{2 e }{h \nu} \sqrt{P_{\mathrm{sig}} P_{\mathrm{LO}}} \frac{d \phi}{d \nu} \delta \nu\left(t\right) + \delta i \left(t\right).
  \label{tdomain_lock}
  \end{equation}
Here, $\phi$ is the frequency-dependent phase shift imparted by the intracavity atomic medium near atomic resonance, $\delta i\left(t\right)$ is the shot noise noise on the photodetector difference signal, $\delta \nu \left(t\right)$ is the system detuning from exact atomic resonance, and $P_{\mathrm{LO(sig)}}$ is the optical power in the LO (signal) pathway. Shot noise will contaminate the resonance condition as 
\begin{equation}
\delta \nu \left(t\right) =  -\delta i \left(t\right) \frac{h \nu}{2 e \sqrt{P_{\mathrm{sig}} P_{\mathrm{LO}}} \frac{d \phi}{d \nu}}. 
\end{equation}
The phase shift near atomic resonance is linear to first order for small frequency deviations, and is given by
\begin{equation}
\frac{d\phi}{d\Delta} =
T_{2}\frac{\mathcal{C}\sigma_z}{\mathcal{C}\sigma_z-1}=
\frac{4T_{2}}{\beta \mathcal{C}} + \mathcal{O}(\mathcal{C}^{-2})\;.
\label{PhaseSlope}
\end{equation}
The shot-noise limited photocurrent noise has a white power spectrum and in the limit of  $P_{\mathrm{LO}} \gg P_{\mathrm{sig}}$, the magnitude is proportional to  $ \frac{e^{2}}{h\nu}P_{\mathrm{LO}} $.
As a consequence, the frequency noise power spectral density of the lock error, $S_{\delta \nu}$, is white. We convert this quantity to conventional laser linewidth (see e.g., \cite{elliot82, stephan2005} and Appendix~\ref{App:Linewidth}) when the system is locked and find that
\begin{equation}
\Delta \nu  = 2 \pi S_{\delta \nu}=
\pi\left(\frac{1}{\mathrm{SNR}\cdot 2\pi\frac{d\phi}{d\Delta}}\right)^2\approx
\frac{\mathcal{C}_{0}}{16\pi \gamma T_{2}^{2}}\beta\;.
\label{eqn:linewidth}
\end{equation}
This is the key result of this paper, as it represents the quantum-limited linewidth, $\Delta \nu$, of a laser stabilized to the nonlinear resonance feature discussed in this work.

It is worth contrasting these results with the ones obtained for a proposed
active laser based on ultra-narrow optical
transitions~\cite{Meiser:SrLaser}.  For that system the linewidth is
given by $\Delta\nu_{\rm laser} = \mathcal{C}_0\gamma/ \pi$.
 The atoms behave more collectively in the case of the laser.  At the peak of laser emission
the collective dipole of the atoms is proportional to $N$,
i.e., $\langle \hat J_+\hat J_-\rangle\propto N^2$, where $\hat
J_-=\hat J_+^\dagger=\sum_{j=1}^N\hat \sigma_-$.  In contrast, for the
passive spectroscopy considered here
\begin{equation}
\langle\hat J_+\hat J_-\rangle
=\frac{N^2}{\mathcal{C}^2}\frac{T_{2} \gamma}{\beta}
\end{equation}
on resonance, $\Delta=0$, {\it i.e.} the effective number of atoms
that participate in the collective dynamics is reduced by a factor
of order $\sqrt{T_{2} \gamma}$. Finally, we note that in the limit where there is no inhomogeneous broadening 
($T_{2} = 2/\gamma$),  Eqn.~(\ref{eqn:linewidth}) reduces to $\Delta\nu = \beta \mathcal{C}_{0}\gamma/ \left(64\pi \right)$. This is, for $\beta$ of order unity, the same scaling as in the laser case.

Table \ref{tab:SpecificSystems} summarizes the stabilization
performance that can be achieved for several atomic species and transitions.   In all
these examples the parameters are chosen such that $\mathcal{C}\approx
100$.  The mode volume of the cavity is $V_{\rm eff}=L\times\pi(100\mu
\mbox{m})^2$, where the length $L$ does not enter the
results. Furthermore, in this locking scheme,
the quantum-limited lock bandwidth (beyond which the signal to noise drops below unity) is given by $\mathrm{BW}_{\mathrm{ql}} = \kappa \mathcal{C}^2n_0\beta$. In all cases considered, this fundamental limitation is well above the kHz range, indicating that the requisite level of laser pre-stabilization is well within current technological capabilities. In several realistic lattice clock systems, we find that laser stabilization can  achieve quantum-limited performance at the mHz level without suffering from thermal noise. Finally, improvements in the coherence time $T_{2}$ of the narrowest transitions yields reciprocal gains in the quantum-limited locked-laser linewidth, underscoring the importance of investigating possible decoherence mechanisms for neutral atom lattice clocks beyond the 1~s time-scale.

In conclusion, we have proposed a laser stabilization technique based
on strongly saturated spectroscopy of narrow optical transitions that
enables linewidths in the 1~mHz range with current experimental
technology.  This technique is not limited by thermal noise and the
fundamental limits of this scheme are below the $1 \mu
\mbox{Hz}$ level.  In the future we plan to study alternative
realizations of this idea including atomic beams and trapped ions.  

We thank J. K. Thompson and J. Cooper for valuable discussions.  This work has been
supported in part by NIST, NSF, DARPA, and ARO.

\appendix

\section{Derivation of locked laser linewidthd}
\label{App:Linewidth}



In this appendix we derive in detail the expression for the quantum noise-limited linewidth, which is presented in Eqn.~12. We begin by considereing the configuration shown in Fig. \ref{fig:AtomCavitySchematic}. The photocurrents of detectors one and two are given by
\begin{align}
      i_{1,2}&=\frac{e\eta_{\mathrm{qe}}}{h\nu}\left[\frac{P_{\mathrm{sig}}}{2}+\frac{P_{\mathrm{LO}}}{2} \pm \sqrt{P_{\mathrm{sig}}P_{\mathrm{LO}}}\cos\left(\Delta \varphi -\phi_{\mathrm{LO}}\right) \right]  \nonumber \\
    &\qquad+ \delta i_{1,2} \left(t\right),
\end{align}
with the ``$+$'' (``$-$'') corresponding to detector one (two). Here, $\Delta \varphi$ is the additional phase shift acquired by the signal beam, $P_{\mathrm{LO}}$ is the power in the local oscillator pathway, $P_{\mathrm{sig}}$ is the power in the signal pathway, $\eta_{\mathrm{qe}}$ is the detector quantum efficiency, and $\delta i_{1,(2)}$ is the stochastically fluctuating component of the photocurrent at detector one (two) due to shot noise. Thus, with the proper choice of LO phase and assuming $\Delta \varphi \ll 1$,
\begin{equation}
i_{\mathrm{diff}}\left(t\right) = i_{1}-i_{2} =\frac{2e\eta_{\mathrm{qe}}}{h\nu}\sqrt{P_{\mathrm{sig}}P_{\mathrm{LO}}}\Delta \varphi +\delta i_{1}\left(t\right) - \delta i_{2}\left(t\right).
\end{equation}
We re-write the term $\delta i_{1}\left(t\right) - \delta i_{2}\left(t\right)$ as $\delta i\left(t\right) \equiv \delta i_{1}\left(t\right) - \delta i_{2}\left(t\right)$.  The time-domain autocorrelation of $\delta i\left(t\right)$ is given by
\begin{align}
\langle \delta i \left(t\right) \delta i \left(t+\tau\right) \rangle &=  \frac{e^{2}\eta_{\mathrm{qe}}}{h\nu}\left[P_{\mathrm{sig}}+P_{\mathrm{LO}}\right]  \delta\left(\tau\right) \nonumber \\
&\simeq \frac{e^{2}\eta_{\mathrm{qe}}}{h\nu} P_{\mathrm{LO}}  \delta\left(\tau\right).
\end{align}
Here, $\delta \left(\tau\right)$ is the Dirac delta function. This corresponds to a two-sided photocurrent noise power spectral density of 
\begin{equation}
S_{i}\left(f\right)= \frac{e^{2}\eta_{\mathrm{qe}}}{h\nu}P_{\mathrm{LO}}.
\end{equation}

\begin{figure}
    \begin{center}
    \includegraphics[scale=.85]{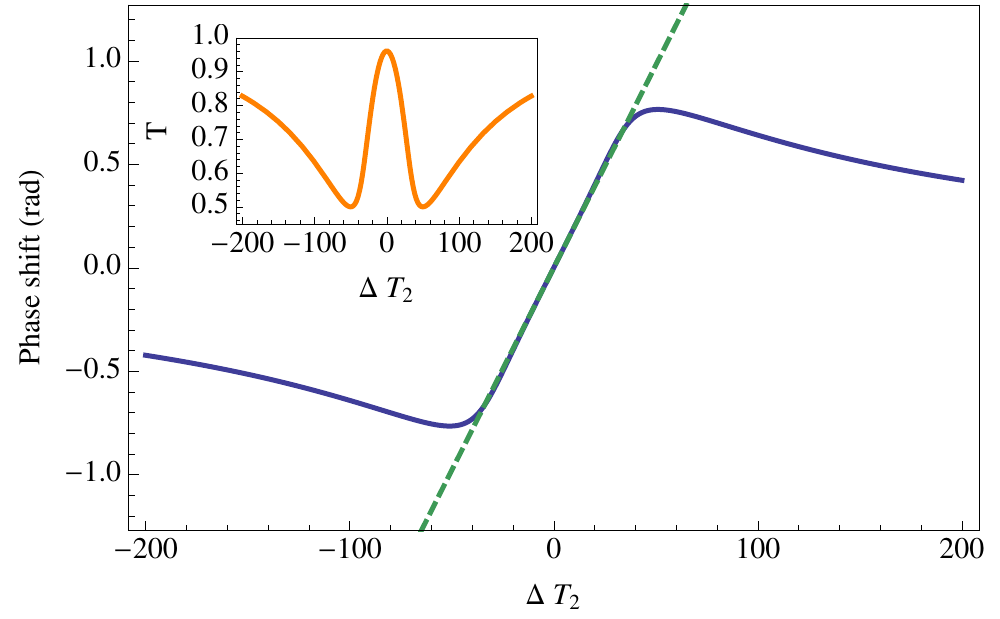}
    \end{center}
    \caption{Phase shift of the transmitted cavity light (with respect to the LO) due to the atomic medium inside the cavity and as a function of laser detuning. Here, $\mathcal{C} = 100$ and $\beta =2$. The dotted line is the linear approximation for the phase near zero detuning. Inset: Cavity transmission curve for the same parameters.}
    \label{Phase}
\end{figure}

The resonance center is observed in this system via the difference photocurrent---namely where the difference photocurrent is equal to zero.  In order to see the effect of the LO shot noise on the lock stability,  one can expand $i_{\mathrm{diff}}$ about zero detuning to linear order of $\Delta \nu$ as
\begin{align}
i_{\mathrm{diff}} & =\frac{2e\eta_{\mathrm{qe}}}{h\nu}\sqrt{P_{\mathrm{sig}}P_{\mathrm{LO}}}\Delta \varphi +\delta i \left(t\right), \nonumber \\  
&\simeq \frac{2e\eta_{\mathrm{qe}}}{h\nu}\sqrt{P_{\mathrm{sig}}P_{\mathrm{LO}}} \frac{\partial \varphi}{\partial \nu}\Delta \nu + \delta i \left(t\right).
\label{idiff}
\end{align}
The validity of making this linear approximation is shown in Fig.~\ref{Phase}, where the complete phase shift of the medium is shown as a function of detuning from resonance along with an analytical solution for the linear phase shift, $\frac{\partial \varphi}{\partial \nu}\Delta \nu$. As long as the laser is close to resonance, the phase is linear to a good approximation. Now we can identify the effect of $\delta i\left(t\right)$ on our ability to determine the line center of the atomic resonance. 

Under locked condition, the DC difference current, $I_{\mathrm{diff}}$, is enforced to be zero via control of the laser frequency. We can thus see that the term $\delta i \left(t\right)$ corrupts the measurement. Namely, our frequency error is given by
\begin{equation} 
\Delta \nu_{\mathrm{err}} \left(t\right) = \frac{\delta i \left( t \right)}{ \frac{2e\eta_{\mathrm{qe}}}{h\nu}\sqrt{P_{\mathrm{sig}}P_{\mathrm{LO}}} \frac{\partial \varphi}{\partial \nu}}.
\label{err_tdom}
\end{equation}
The denominator came directly from Eqn.~\ref{idiff}. 


When locked to the cavity-atom resonance, we assume the laser has an electric field given by
\begin{equation}
E(t) = E_0 e^{i 2 \pi \nu_0 t + \delta \phi(t)}.
\end{equation}
Here the phase error, $ \delta \phi(t)$, is related to $\Delta \nu_{\mathrm{err}} \left(t\right)$ by  
\begin{equation}
\frac{d \delta \phi}{ d t } \equiv \Delta \nu_{\mathrm{err}} \left(t\right). 
\label{derivative}
\end{equation}

In order to go from this time-domain expression to the frequency domain via the Wiener-Khinchin theorem (following the general approach presented in Ch.~3 of \cite{Riehle2004}), we compute the autocorrelation of the field amplitude, $R_{E} \left(\tau\right)$, given by
\begin{align}
 R_{E} \left(\tau\right) &=  \langle E(t) E^{\star}(t +\tau) \rangle \nonumber \\
 &= \left|E_0\right|^{2}  e^{i 2\pi \nu_0 \tau} \langle e^{i\left(\delta \phi (t) - \delta \phi(t+\tau)\right)}\rangle.
 \end{align}
An application of the gaussian moment theorem gives 
\begin{equation}
\langle e^{i\left(\delta \phi (t) - \delta \phi(t+\tau)\right)}\rangle =\mathrm{exp} \{- \langle \left[ \delta \phi (t) - \delta \phi(t+\tau)\right]^{2}\rangle/2\}.
 \end{equation}
We can re-write the expectation value as
\begin{align}
 \langle \left[ \delta \phi (t) - \delta \phi(t+\tau)\right]^{2}\rangle &= 2 \langle \left[ \delta \phi(\tau)\right]^{2}\rangle - 2 \langle \left[ \delta \phi (t) \delta \phi(t+\tau)\right]\rangle \nonumber \\
& =  2\left[R_\phi \left(0\right) - R_\phi \left(\tau\right)\right].  
\end{align} 
It is then a direct consequence of the Wiener-Khinchin theorem that 
\begin{equation}
 \langle \left[ \delta \phi (t) - \delta \phi(t+\tau)\right]^{2}\rangle  = 2 \left[ \int_{-\infty}^{\infty} S_{ \delta \phi} \left(f\right) \left(1 - e^{i 2 \pi f \tau} \right) df \right],
 \end{equation}
 where $S_{\delta \phi} \left(f\right)$ is the two-sided phase fluctuation power spectral density for $\delta \phi (t)$.
However, we can easily relate $S_{\delta \phi}$ to $S_{\Delta \nu}$ (the two-sided frequency deviation power spectral density) by Eqn.~\ref{derivative}, such that 
\begin{equation}
 \langle \left[ \delta \phi (t) - \delta \phi(t+\tau)\right]^{2}\rangle  = 2 \left[ \int_{-\infty}^{\infty} \frac{S_{ \Delta \nu_{\mathrm{err}}} \left(f\right)}{f^{2}} \left(1 - e^{i 2 \pi f \tau} \right) df \right].
 \label{integral_eq}
 \end{equation}

Applying the Wiener-Khinchin theorem to Eqn.~\ref{err_tdom}, we have 
\begin{equation}
S_{ \Delta \nu_{\mathrm{err}}} =  \frac{h \nu}{4 \eta_{\mathrm{qe}} P_{\mathrm{sig}}  \left(\frac{\partial \varphi}{\partial \nu}\right)^{2}} 
\end{equation}
We can therefore re-write Eqn.~\ref{integral_eq} as 
\begin{align}
\langle \left[ \delta \phi (t) - \delta \phi(t+\tau)\right]^{2}\rangle  &=  \int_{0}^{\infty} \frac{h_{0}}{f^{2}} \left[1 - \cos \left(2 \pi f \tau\right) \right] df \nonumber \\
& = h_{0} \pi^{2} \tau,
\end{align}
with $h_0$ given by
\begin{equation}
h_0 = \frac{h \nu}{\eta_{\mathrm{qe}} P_{\mathrm{sig}}  \left(\frac{\partial \varphi}{\partial \nu}\right)^{2}} .
\end{equation}
Now we have an expression for the electric field autocorrelation, namely
\begin{equation}
R_{E} \left(\tau\right) = \left|E_0\right|^{2}  e^{i 2\pi \nu_0 \tau} e^{-h_0 \pi^{2} \tau/2}.
\end{equation}
We apply the Wiener-Khinchin theorem to this expression and obtain a Lorentzian profile for the laser optical power, with frequency full width half maximum, $\Delta \nu _{\mathrm{FWHM}}$, given by 
\begin{equation}
\Delta \nu _{\mathrm{FWHM}} = \frac{\pi h_{0}}{2} =   \frac{\pi h \nu}{2\eta_{\mathrm{qe}} P_{\mathrm{sig}}  \left(\frac{\partial \varphi}{\partial \nu}\right)^{2}}.
\end{equation}
We combine this with the results of Eqns.~8 and 11 of the main text, and obtain the result presented in Eqn.~12, in the limit of unity detector quantum efficiency, namely
\begin{equation}
\Delta \nu   \approx
\frac{\mathcal{C}_{0}}{16\pi \gamma T_{2}^{2}}\beta\;.
\label{eqn:linewidth}
\end{equation}

\section{Line-pulling effects due to cavity-laser detuning}
\label{LinePull}
In order to derive the line-pulling due to an imperfect lock between the cavity and probe laser, we make use of the full optical bistability equation that describes the input/output dynamics of the system \cite{gripp1997:evolution},
\begin{equation}
y = x\left(1 + \mathcal{C} \frac{1-i T_{2}\Delta}{1 + |x|^{2} + \left(T_{2} \Delta\right)^{2}} + i \theta\right). \label{InOut1}
\end{equation}
The parameter $x$ is related to $\langle a \rangle$ by  $x =  \langle a \rangle/\sqrt{n_{0}}$, $y$ is given by $\eta / \left(\kappa \sqrt{n_0} \right)$ ($\left|y\right|^{2} = I_{\mathrm{in}}$), $\Delta$ and $T_{2}$ are the same as given in the text, and the parameter $\theta$ is the cavity-laser detuning in units of $\kappa$, namely $\theta = \left(\omega_{c} -\omega_{l}\right)/\kappa$.  In the text, $\theta$ was assumed to be negligibly small. Here we quantify this statement.

If we assume that we are near resonance in the nonlinear, strongly saturated regime ($\beta >1$, $\mathcal{C} \gg 1$), then $\left|y\right|^{2} \simeq \left|x\right|^{2} = \beta \mathcal{C}^{2}/4$. If $T_{2} \Delta\ll \mathcal{C}$, then we can expand Eqn.~\ref{InOut1} such that 
\begin{equation}
y \simeq x\left[1 + \frac{4}{\mathcal{C}\beta} \left(1-i T_{2} \Delta\right)+ i \theta\right]. \label{InOut2}
\end{equation} 
Therefore, the phase shift of the transmitted light is given by
\begin{equation}
\Delta \phi = \mathrm{Arg} \left[x/y\right] \simeq  \frac{4 T_{2} \Delta}{\mathcal{C}  \beta} - \theta . 
\label{simple_phase}
\end{equation}

From Eqn.~\ref{simple_phase}, it can be seen that for a given cavity-laser detuning, the lock center frequency shift, $\Delta \nu_{\mathrm{laser}}$, is given by
\begin{equation}
\Delta \nu_{\mathrm{laser}} = \frac{\mathcal{C} \beta}{8 \pi T_{2} } \left(\frac{\omega_{c} -\omega_{l}}{\kappa}\right). 
\end{equation}
Cavity lock precisions of $>10^{4}$ are routinely achieved in the laboratory. This implies that
\begin{equation}
\left(\frac{\omega_{c} -\omega_{l}}{\kappa}\right)\lesssim10^{-4}.
\end{equation}
For typical parameters considered in the main text, namely $\mathcal{C} = 100$, $\beta=2$, and $T_{2} =1$~s, this implies that the cavity pulling effect is below the 1~mHz level. Longer $T_{2}$ times will further suppress this effect.


%

\end{document}